# Super-concentrated alkali hydroxide electrolytes for rechargeable Zn batteries

Yilin Ma,[a,1] Jiajia Huang,[b,1] Shengyong Gao,[a] Liangyu Li,[c] Zhibin Yi,[c] Diwen Xiao,[c] Cheuk Kai Kevin Chan,[c] Ding Pan,[*b, a, d] and Qing Chen,[*c, a]

a.  Department of Chemistry, The Hong Kong University of Science and Technology, Clear Water Bay, Hong Kong, P. R. China.
b.  Department of Physics, The Hong Kong University of Science and Technology, Clear Water Bay, Hong Kong, P. R. China.
c.  Department of Mechanical and Aerospace Engineering, The Hong Kong University of Science and Technology, Clear Water Bay, Hong Kong, P. R. China.
d.  HKUST Shenzhen-Hong Kong Collaborative Innovation Research Institute, Shenzhen, China

[1] These authors contributed equally.
* Corresponding authors: Ding Pan, Email: dingpan@ust.hk;
Qing Chen, Email: chenqing@ust.hk

**Abstract:** Rechargeable Zn batteries offer safe, inexpensive energy storage, but when deeply discharged to compete with lithium-ion batteries, they are plagued by parasitic reactions at the Zn anodes. We apply super-concentrated alkaline electrolytes to suppress two key parasitic reactions, hydrogen evolution and ZnO passivation. An electrolyte with 15 M KOH displays a broad electrochemical window (>2.5 V on Au), a high ZnO solubility (>1.5 M), and an exceptionally high ionic conductivity (>0.27 S/cm at 25 °C). Spectroscopies and ab-initio molecular dynamics simulation suggest $K^+$-$OH^-$ pairs and a tightened water network to underpin the stability. The simulation further reveals unique triggered proton hopping that offsets the lack of 'water wires' to sustain the conductivity. Low hydrogen evolution, confirmed via online mass spectroscopy, and slow passivation enable a NiOOH||Zn battery to deliver a cumulative capacity of 8.4 Ah/cm$^2$ and a Zn-air battery to last for over 110 hours.

Super-concentrated aqueous electrolytes bring exciting opportunities for chemical engineering of electrode/electrolyte interfaces for safe, durable rechargeable batteries.[1,2] The salt concentrations of these electrolytes are an order of magnitude higher than that of a conventional one, suppressing water splitting while boosting desirable salt decomposition reactions. The most prominent case is the 'water-in-salt' electrolyte comprising ~20 M lithium bis(trifluoromethane sulfonyl)imide (LiTFSI), whose unique solution structure broadens the electrochemical window of water to over 3 V to survive in lithium-ion batteries (LIBs).[3] Similar designs underlie hydrate-melt and molecular crowding electrolytes, all of which strive to replace flammable organic solvents with water for safer LIBs.[4,5]

The impacts of these electrolytes have lately been brought on LIB alternatives, including rechargeable Zn batteries.[6,7,8] Touted as cheaper, safer alternatives to LIBs, the performance of aqueous Zn batteries is plagued by hydrogen evolution reaction (HER), as the equilibrium potential of $Zn^{2+}$/Zn is far below that of a reversible hydrogen electrode at all pH's, causing losses of Coulombic efficiency (CE), electrolyte volume, and battery capacity.[9-14] The issue is lessened with super-concentrated aqueous electrolytes, including 30 m $ZnCl_2$, 1 m $Zn(TFSI)_2$ + 20 m LiTFSI, 4 m $Zn(CF_3SO_3)_2$ + 2 m $LiClO_4$, 0.5 m $ZnSO_4$ + 21 m LiTFSI and acetate (Ac) based electrolytes.[7,8,15-20] The intriguing properties and outstanding performance motivate extensive spectroscopic and computational investigations into the unprecedented solution structures, revealing the roles of water-lean coordination[7] and ion aggregates[20], although questions remain in crucial structure-property relationships[21].

Unlike these near-neutral electrolytes, alkaline electrolytes have not been explored to a super high concentration for Zn batteries. As one of the oldest batteries, alkaline Zn batteries face no smaller challenge of HER, which is thermodynamically more favored at higher pH. The batteries traditionally run with a concentration of 4 – 6 M KOH for the highest ionic conductivity.[22,23] There have been limited attempts to go for higher concentrations. Early work by Cairns and others argued that too high a KOH concentration led to a high solubility of zincate ions (in the form of $Zn(OH)_4^{2-}$), which diffuse to redistribute Zn mass and cause shape changes.[24,25] However, a recent push towards a high depth of discharge (DoD) of Zn anodes, necessary for the most mature NiOOH||Zn cells to reach LIB-level energy density, has shown the benefit of higher KOH concentrations.[23] Along this line, Park et al. screened a variety of electrolytes for Zn foam anodes and identified 6 M KOH with 1 M LiOH and a $Ca(OH)_2$ suspension as one of the best.[26] Lim et al. investigated the effect of zincate saturation in the electrolyte and attained the best cycling performance at 7 M KOH, although they blamed an even higher concentration for the redistribution of Zn mass.[27] We showed that a nanoporous Zn anode developed in our group worked the best at 40% DoD in 6 M KOH and 1 M LiOH, and the KOH concentration had to be raised to 9 M for stable cycling at 60% DoD.[28] Chen et al. recently exploited KOH concentrations up to 22 M (although the concentrated electrolytes beyond the solubility are metastable) in Zn-air batteries, but the focus was on the freezing properties at low temperatures, for which they claimed 7 M was optimal.[29]

Here, we demonstrate the use of a super-concentrated alkaline electrolyte (SCAE) for extending the cycle lives of alkaline Zn batteries. We combine electrochemical methods with spectroscopies and ab initio molecular dynamics (AIMD) simulations to elucidate the solution structures and a novel proton hopping mechanism underlying the unique properties of the electrolytes, including the wide electrochemical window and the high ionic conductivity. These properties are retained as we engineer the electrolytes to work in NiOOH||Zn and Zn-air batteries to address hydrogen evolution and ZnO passivation for stable cycle lives without compromising the energy densities.

*Electrochemical stability and conductivity*

The hypotheses underlying the design of SCAEs are explained in Fig. 1a, which concerns two key properties, the electrochemical stability of water (characterized by the electrochemical window, $E_w$) and the ionic conductivity ($\kappa$). In dilute or moderately concentrated aqueous electrolytes, not only do water molecules have a high activity, they may also transfer electrons to metal cations, thus weakening the O–H bonds for rapid water splitting.[10, 30-34] On the contrary, a super-concentrated electrolyte reduces the number of water per liter of electrolyte and limits its presence in the solvation shell of the cations. A common drawback is a high viscosity ($\eta$) and thus a low $\kappa$ according to the Nernst-Einstein equation. However, hydroxides are exceptionally mobile ions thanks to the structural diffusion via proton hopping among hydroxides and water (i.e., the Grothuss mechanism). While the high concentration lowers the number of water available for such structural diffusion, the rise of hydroxide molarity may retain a strong percolation of proton hopping paths. Therefore, SCAEs offer an opportunity to decouple $\kappa$ from $\eta$ or at least retain high $\kappa$ even with high $\eta$.

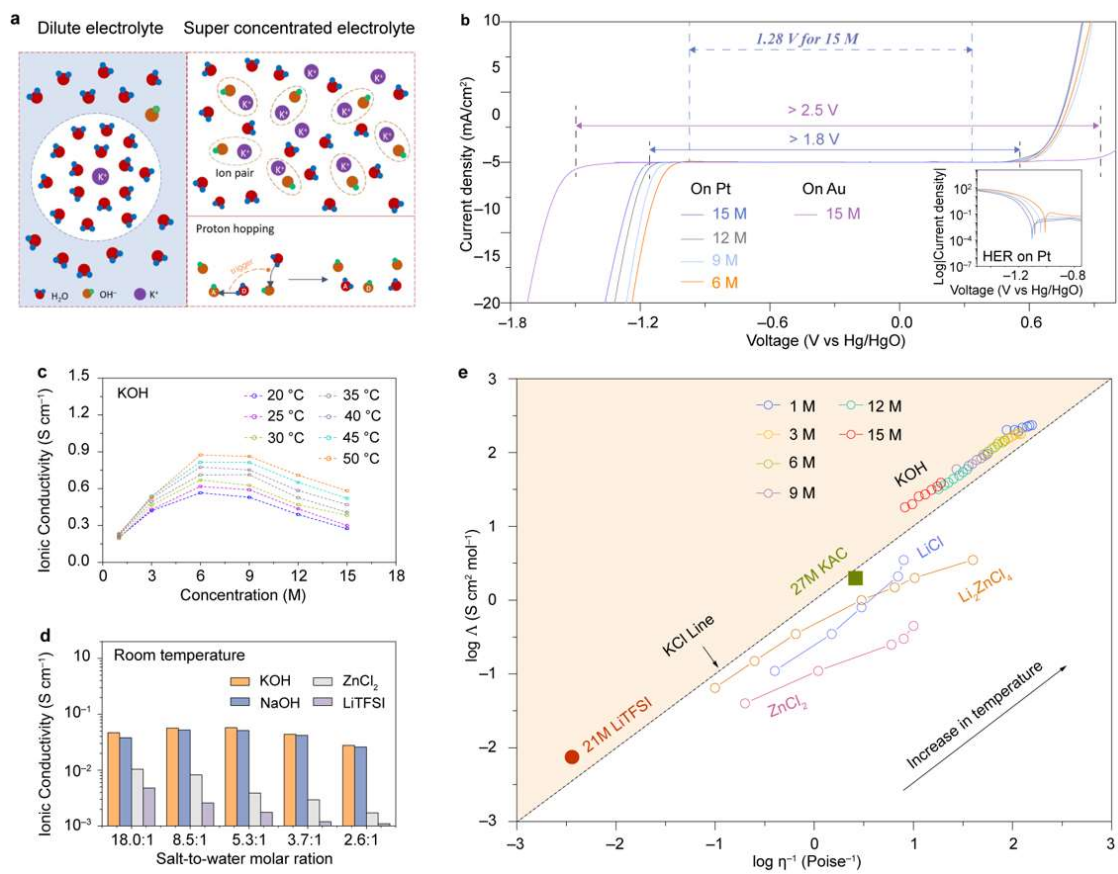

**Fig. 1 | Properties of SCAEs. a**, An illustration of how solution structures at different concentrations can lead to different properties, **b**, Linear sweep voltammetry performed in 6 – 15 M KOH electrolytes on Pt and Au electrodes. The insert shows Tafel curves on Pt in the region of HER. The dashed, double-arrow line shows the calculation of $E_w$ with $C_{KOH}$ =15 based on reported thermodynamic properties, highlighting the significant contribution of the kinetics. **c**, Ionic conductivity vs. $C_{KOH}$ at a temperature range of 20 – 50 °C (see Table S2). **d**, Ionic conductivity vs. salt/water molar ratio for four types of electrolytes (KOH, NaOH, $ZnCl_2$, LiTFSI), (see Table S3). **e**, A Walden plot for various super-concentrated aqueous electrolytes. The perfector $C_\Lambda$ is 1, 3, 6, 9, 12 and 15 for 1, 3, 6, 9, 12 and 15 M KOH electrolytes (see Fig. S3 and Table S4 for the viscosity value), 1 for LiCl·$3H_2O$, 3 for $ZnCl_2$·$3H_2O$ and 5/3 for $Li_2ZnCl_4$·$9H_2O$[35], 21 for 21 M LiTFSI electrolyte[36] and 27 for 27 M electrolyte[37].

We first confirm the high electrochemical stability of SCAEs via linear sweep voltammetry (LSV). We focus on KOH as it is the most conductive among common alkali hydroxides. We select electrolytes of a KOH concentration ($C_{KOH}$) of 6, 9, 12, and 15 M, respectively (see Table S1 for details), in which the lowest is common in batteries and the highest approaching the solubility of KOH in water[38] (Fig. S1). We categorize the two of $C_{KOH}$ = 12 and 15 M as SCAEs. Pt or Au serves as the working electrode. The electrochemical window $E_w$ is defined as the difference between the voltages at which 0.05 mA/cm$^2$ of anodic and cathodic currents are recorded respectively at a scan rate of 5 mV/s. Even on the catalytic Pt electrode, $E_w$ increases substantially with $C_{KOH}$. In 15 M KOH, it approaches 1.8 V, a value much larger than what is predicted based on the water activity decrease (shown as dashed double arrows in Fig. 1b, with calculation details in Fig. S2). It underlines the importance of kinetics, more significant on the cathodic side when we plot the current at a logarithmic scale, in line with the suppression of HER observed for increasing pH.[39] $E_w$ expands further to 2.5 V on the more inert Au electrode (Fig. 1b).[40] The stability is comparable with other super-concentrated electrolytes.[1-6, 20]

The electrolytes display high ionic conductivities as we postulated. Our measurements are performed with a conductivity meter with electrolytes of $C_{KOH}$ = 1, 3, 6, 9, 12, and 15 M, respectively at 20 to 50 °C. The conductivity peaks at 0.567 S/cm at $C_{KOH}$ = 6 M at room temperature, consistent with the literature[29, 41] (Fig. 1c). What is more remarkable is that even as $\eta$ hikes to ~10 mPa·s at $C_{KOH}$ = 15 M, $\kappa$ remains at 0.277 S/cm, nearly an order of magnitude higher than other electrolytes at similar salt-to-water ratios (Fig. 1d). Similarly high conductivities are measured for SCAEs with NaOH, highlighting the role of the hydroxide ions. The conductivity stands out among super-concentrated electrolytes in a Walden plot, where we convert $\kappa$ to molar conductivity ($\Lambda$) and plot it against $1/\eta$ on a double-logarithmic scale. Although it is debatable whether such a plot can quantify the ionicity of an electrolyte[42], we can still gain insights into the conduction mechanism. The plot includes conductivities measured under different temperatures at different concentrations. A dashed diagonal line passes through the values of 1 M KCl aqueous solution, providing a routine reference, which separates KOH from other salts. At a similar viscosity, the vertical separation between SCAEs and the others is about one order of magnitude (Fig. 1e), similar to the difference in diffusivity between OH$^-$ and other anions. All measured values for KOH seem to aggregate onto a single line, suggesting its full dissociation even close to the saturation, consistent with previous reports.[43] The results also confirm the rapid conduction independent of the concentration, likely via the structural diffusion.

*Solution structures*

To understand the solution structure behind the intriguing properties, we employ spectroscopies and simulations. $^1$H nuclear magnetic resonance (NMR) spectra reveal significant deshielding as $C_{KOH}$ increases (Fig. 2a), indicating changes in the hydrogen-bond network. The $^1$H peak shifts from 4.31 ppm in H$_2$O to 4.74 ppm in 15 M KOH. The shift to low field, the deshielding of the hydrogen nuclei due to a lower electronic density, suggests extensive changes to the hydrogen-bond network.[11] The interpretation is supported by a control experiment performed on 1 M KCl (Fig. S4), in which the $^1$H peak shifts to a high field, so the dishielding in KOH is indeed due to OH$^-$. In $^{39}$K NMR spectra, the $^{39}$K peak shifts positively with $C_{KOH}$ (Fig. 2b). A possible explanation is that although higher $C_{KOH}$ brings more OH$^-$ into the solvation sheath of K$^+$ to shield it, the solvation sheath may be shared by adjacent K$^+$ ions given insufficient water molecules[44].

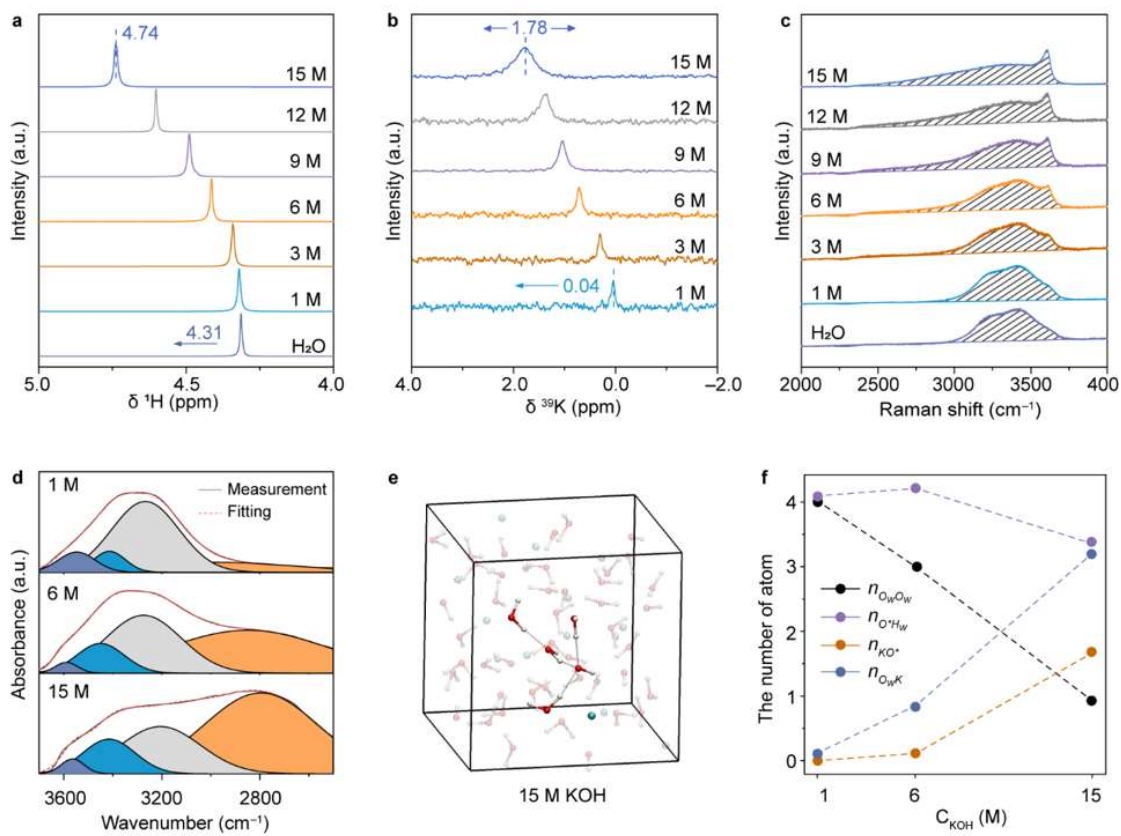

**Fig. 2 | Spectroscopies and simulations of the SCAE. a**, Chemical shifts of $^1$H nuclei in the KOH electrolytes and pure water. **b**, Chemical shifts of $^{39}$K nuclei in the KOH electrolytes. **c**, Raman spectra of the KOH electrolytes and pure water. **d,** The deconvolution of the OH stretching peaks for the FTIR spectra of 1, 6 and 15 M KOH respectively. **e,** A snapshot of a 15 M KOH solution from the AIMD simulation. K, O and H atoms are in cyan, red, and white, respectively. Hydrogen bonds, defined according to Ref $^{49}$ are highlighted as gray dashed lines. **f,** A summary of the coordination number for the central atom vs. $C_{KOH}$.

Raman and Fourier transform infrared (FTIR) spectra reveal more structural details in SCAEs. In Fig. 2c, Raman spectroscopy captures the emergence of a sharp peak at 3606 cm$^{-1}$, which originates from the interaction between K$^+$ and OH$^-$ as seen in the spectrum of a KOH solid (Fig. S5). It suggests that even though the two ions in SCAEs remain dissociated as seen in our conductivity measurements and the literature, they may form ion pairs for the lack of water. Meanwhile, a broadening inhomogeneous feature from 2450 cm$^{-1}$ to 3360 cm$^{-1}$ signals new modes of O–H stretching vibration.[45] FTIR spectroscopy (Fig. S6) shows a decay of the broad band at 3000 – 3700 cm$^{-1}$ that belongs to $H_2O$[41] and a rise of the band near 2900 cm$^{-1}$ that belongs to OH$^-$ (see comparisons in Fig. S7) as $C_{KOH}$ increases.[46,47] Consistent with the Raman spectroscopy, it reveals a peak near 3600 cm$^{-1}$ in 15 M KOH due to the cation-anion interaction, corroborated by spectra of other solutions (Fig. S8). We deconvoluted the OH stretching peaks into four components: weak hydrogen-bond water at 3530 cm$^{-1}$, asymmetric hydrogen-bond water molecules at 3400 cm$^{-1}$ (i.e., water-cation interaction), symmetric hydrogen-bond water molecules at 3260 cm$^{-1}$ (water-water interaction), and the OH$^-$-bonded water near 2800 cm$^{-1}$ (Fig. 2d). In 15 M KOH, the emergence of the lower-frequency broad continuum (at 2800 cm$^{-1}$) indicates the destruction of the water structure at lower concentrations (Fig. S9).[46-48]

To gain atomistic insights into the solution structure of SCAE, we perform AIMD simulations (details in Table S5). We intend to compare SCAEs to lower concentration solutions in previous computational[50-52] and experimental[48,53] studies for further insights. A snapshot of the simulated structure at $C_{KOH}$ = 15 M (Fig. 2e) shows that water molecules lose the tetrahedral-like network[54] at lower $C_{KOH}$ (in black in Fig. 2f and Fig. S10) and form a more compact structure, consistent with previous reports.[48] It is consistent with quantitative analyses (see details in Fig. S11) based on radial distribution functions (RDFs), denoted as $g_{xy}(r)$, providing the probability of finding an atom $y$ at a specific distance $r$ from $x$. Integrating a RDF yields a cumulative distribution function $N(r)$, whose value at the first minimum of the RDF provides the coordination number for the central atom. We denote the oxygen and hydrogen atoms in OH$^-$ as O$^*$ and H$^*$, and those in water as O$_w$ and H$_w$ respectively.

As the $C_{KOH}$ increases from 1 M to 15 M, the coordination number of O$^*$ derived from the RDF decreases significantly from 4.1 to 3.4 (in violet in Fig. 2f). This observation suggests that the solvation structure of OH⁻ in dilute KOH solutions primarily consists of $OH^-(H_2O)_4$, whereas at $C_{KOH}$ = 15 M, a new structure $OH^-(H_2O)_3$ emerges. At the low concentrations (1 M and 6 M), O* tends to accept four hydrogen bonds, while at the high concentration, O* is more inclined to accept three hydrogen bonds(Fig. S12).

The simulation confirms K⁺−OH⁻ ion pairs in the SCAE. At $C_{KOH}$= 15 M, the $g_{KO*}$ (Fig. S13) displays a prominent peak at 2.7 Å and a coordination number of 2 at 3.5 Å, corresponding to K⁺−OH⁻ ion pairs not observed at the lower concentrations (summarized in orange in Fig. 2f). Additionally, the outward migration of the second $g_{KOw}$ peak (Fig. S14) helps to compensate for the reduced solvent dielectric constant within highly alkaline environments[55], supporting our interpretation of the NMR spectra. It is also important to acknowledge, as proposed by Shao et al.,[56] that the formation of ion pairs does not necessarily decrease electrolyte conductivity. Meanwhile, the replacement of water by OH⁻ in the solvation shell of K⁺ can reduce the influence of K⁺ on water that is believed to exacerbate HER. Another intriguing observation is that the coordination number of $g_{OwK}$ increases from 0.1 at 1 M to 3.2 at 15 M (in blue in Fig. 2f), meaning that more K⁺ ions share water molecules. It also weakens water-water interactions, leading to a distinct hydrogen-bond network revealed earlier by both the spectroscopies.

*Proton hopping*

The high ionic conductivity of SCAEs roots in proton hopping. While a low water concentration is considered to limit the rate of pre-solvation, [44, 51, 57] a crucial step in the Grotthuss process, our observation of the slow decline in ionic conductivity with $C_{KOH}$ motivates us to further scrutinize proton hopping. We define a proton hopping coordinate $\delta$ as shown in Fig. 3a, following the method proposed by Ref. 58. When $\delta = 0$ Å, a proton is halfway between two oxygens (the smallest $|\delta|$ is chosen among multiple values arising from may OH$^-$···HOH pairs of one OH$^-$). A small value of $|\delta|$ indicates that the proton is in a hopping state, undergoing transfer between two oxygen atoms. Conversely, a larger value of $|\delta|$ suggests that the proton is well-localized within the first solvation shell of OH$^-$. To distinguish these states, we calculated the free energy as a function of $|\delta|$,

$$F(|\delta|) = -k_B T \ln P(|\delta|) \tag{1}$$

where $P(|\delta|)$ denotes the probability distribution function of $|\delta|$. Fig. 3a shows a maximum at $\delta = 0$ Å and a local minimum near $\delta = 0.42$ Å with $C_{KOH} = 1$ and 6 M and 0.48 Å $C_{KOH} = 15$ M respectively. The difference between the minimum and the maximum is the free energy barrier (1.72, 1.76, and 2.00 kcal/mol at $C_{KOH} = 1$, 6 and 15 M, respectively). We consider that the proton hopping is at the initial or final state when $|\delta|$ is larger than 0.5 Å, and is at the transition state when $|\delta|$ is smaller than 0.1 Å. A comparison of RDFs between these two states reveals solution structural changes during proton hopping.

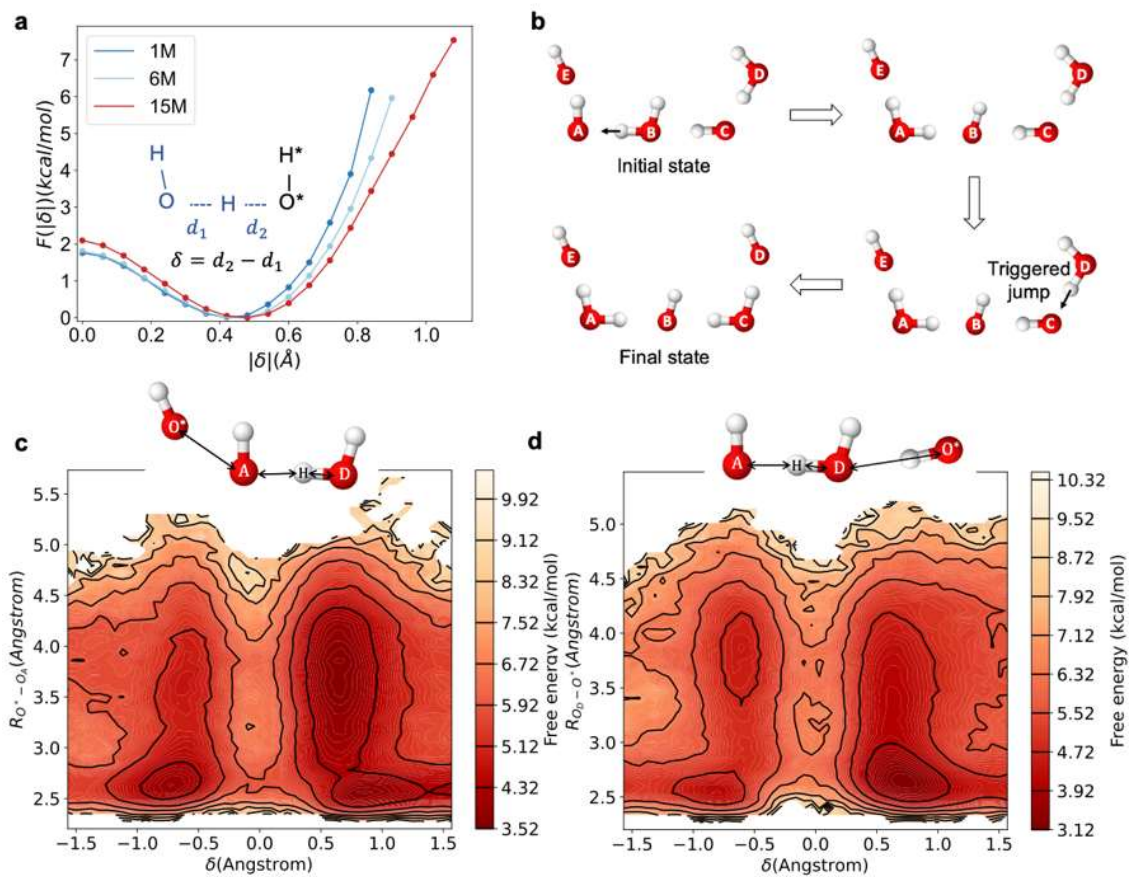

**Fig. 3 | Triggered proton hopping. a,** The free energy function of the proton hopping coordinate $|\delta|$. **b,** Steps of the hopping mechanism. The arrow shows the direction of proton movement. Step 1: a proton hops from H$_2$O($O_B$) to OH$^-$($O_A$). Step 2: another proton is triggered to hop. **c,** The iso-contours of the free energy of $R_{O_A-O^*}$ for proton hopping between OH$^-$($O_A$) and H$_2$O($O_B$) in the 15 M KOH solution. $R_{O_A-O^*}$ represents the distance between $O_A$ from the accepting OH$^-$ and $O^*$ from the OH$^-$ ion nearest to $O_A$. $\delta$ is the proton hopping coordinate as defined earlier ($d_2 - d_1$). **d,** The iso-contours of the free energy of $R_{O_B-O^*}$ for proton hopping between OH$^-$ ($O_A$) and H$_2$O($O_B$) in the 15 M KOH solution. $R_{O_B-O^*}$ represents the distance between $O_B$ from the donating H$_2$O and O$^*$ from the OH$^-$ ion (not the accepting OH$^-$) nearest to $O_B$.

From the solution structural changes, a rather surprising observation pertains to the stable coordination number of water in SCAE. While g$_{OwH}$ (Fig. S15) shows that a tetrahedral arrangement prevails at low $C_{KOH}$, the coordinate number of water at $C_{KOH}$ = 15 remains at 3 throughout the proton hopping. It differs from a previous conclusion by Zhu and Tuckerman[51] that the scarcity of water in concentrated KOH solutions necessitates the acquisition of an additional hydrogen bond in the solvation shell of water in the proton hopping, responsible for the lower conductivity. They observed that both O$^*$ and O$_w$ had a similar local structure with an approximate coordination number of 4 at the transition state, which could be due to the semi-local exchange-correlation functional BLYP used in their study, resulting in an overstructured hydrogen bond network.[59] Our observations instead suggest that the pre-solvation process likely involves OH$^-$ ions directly adopting the solvation structure of water at the corresponding concentrations. Therefore, the rate of proton hopping is not significantly limited by the scarcity of water molecules. Instead, the metastable and the transition states are structurally similar across $C_{KOH}$, which partly explains the earlier similar energy barriers of proton hopping.

We can gain more insights into the energy barrier through free energy maps, which reveals a mechanism of triggered proton hopping in SCAE (Fig. 3b). The mechanism is illustrated in Fig. 3b and a Supplementary Movie S1, where two proton hopping events occur without a proton wire, based on an analysis of all OH$^-(O_A)\cdots$H$_2$O($O_B$) pairs that undergo proton exchange for which we calculated the free energy maps. In Fig. 3c, before the proton hopping (δ > 0, the starting arrangement in Fig. 3b), there are two valleys corresponding to the first ($O_E$) and second ($O_C$) solvation O$^*$ shells of OH$^-$ ($O_A$). After the proton hopping (δ < 0), the valley at $R_{O_A-O^*}$~3.4 Å becomes shallower than the valley at $R_{O_A-O^*}$~2.6 Å, triggering a proton to hop to $O_C$ in an OH$^-$ ion in the second solvation shell. This triggered proton hopping is further confirmed in Fig. 3d, where the valley at $R_{O_B-O^*}$~2.6 Å also becomes shallower relative to the valley at $R_{O_B-O^*}$~3.4 Å after the proton hopping from $O_B$ to $O_A$ (δ < 0), which indicates that $O_C$, the OH⁻ ions in the first solvation shell of $O_B$ convert into H₂O. At 15 M, the proton hopping can begin with the OH⁻···OH⁻($O_A$)···H₂O($O_B$)···OH⁻($O_C$) complex, where the unbalanced distribution of OH⁻ creates an electrostatic force that attracts the proton. Once the proton is successfully transferred to $O_A$, the OH⁻($O_B$)···OH⁻($O_C$)···H₂O complex is formed, triggering another proton hopping event in which $O_C$ accepts the proton. These two jumps help establish a more balanced distribution of OH⁻ ions throughout the overall configuration. This mechanism is absent at 6 M (Fig. S16), and it differs from the concerted or stepwise proton hopping that is commonly studied in low-concentration solutions[60,61], where a continuous water wire acts as the proton carrier. Overall, our calculations suggest that although the coordination numbers of water molecules and OH⁻ ions in the SCAE are lower, it still involves the pre-solvation process to facilitate proton transfer. The scarcity of water is offset by more frequent hopping (Fig. S17) and a new mechanism of triggered proton hopping (Fig. 3b), maintaining the high ionic conductivity.

*Suppressing side reactions*

We now assess the effectiveness of SCAEs in suppressing two key side reactions in high-DoD Zn batteries, HER and anode passivation. The large value of $E_w$ measured earlier does not warrant less HER, which depends also on the reduction potential of zincate. While thermodynamics predicts the reduction voltage of zincate to descend more rapidly with increasing alkalinity than the voltage of HER, the sluggish kinetics of HER helps mitigate this side reaction, evident from LSV performed with 0.1 M ZnO dissolved in SCAEs, where the reduction of zincate precedes HER (Fig. S18). We further confirm the suppression of HER through differential electrochemical mass spectrometry (DEMS). The electrolytes are assembled into coin cells equipped with gas passages, from which hydrogen gas released during Zn deposition can be sensed (Fig. S19). We do not add any ZnO into the electrolytes to ensure that the electrode/electrolyte interface is free of oxide, so that the measurement would yield the maximum rates of HER intrinsic of the electrolytes. Once the cell begins to charge (at the 180$^{th}$ minute), the rate spikes for $C_{KOH}$ = 6 M but not the others (Fig. 4a). For $C_{KOH}$ = 15 M, the rate of hydrogen evolution remains at a level similar to that in discharging. We can quantify the effectiveness of HER suppression with a charging efficiency loss, calculated as the ratio between the charge going to HER (estimated via the Faraday's law) and the overall charge input. With $C_{KOH}$ = 6 M, the efficiency loss is as high as 10.8%, which drops to 1.6% with $C_{KOH}$ = 15 M (Fig. 4b).

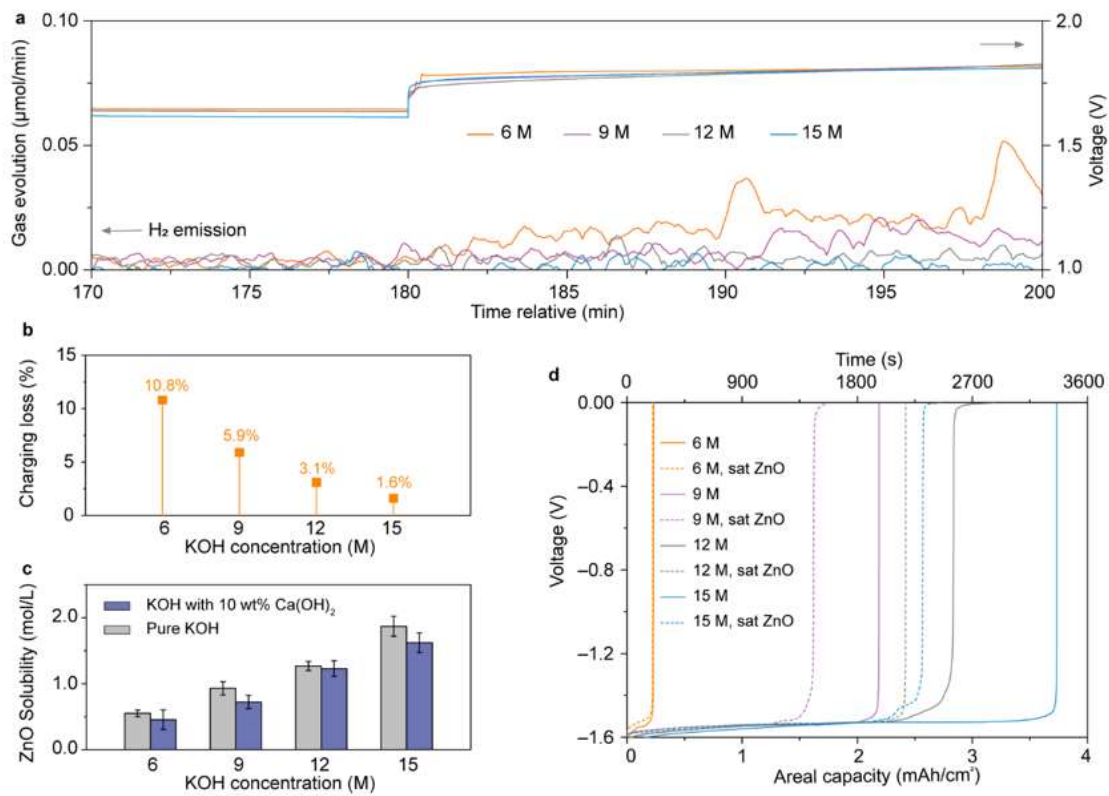

**Fig. 4 | Suppression of HER and passivation in SCAE. a,** Hydrogen evolution rate (bottom) measured via DEMS connected to a Zn||NiOOH cell with electrolyte of labeled $C_{KOH}$, which undergoes galvanostatic discharging and charging as shown by the voltage-time curves (top). **b,** Charging efficiency losses estimated from the charging processes in **a**. **c,** ZnO solubility in

electrolytes of different values of $C_{KOH}$ with and without additional 10wt% Ca(OH)$_2$. **d,** Chronopotentiometric curves of Zn electrodes in electrolytes of different $C_{KOH}$ values with no or saturated ZnO.

The issue of passivation is tied to the solubility of zincate. Upon discharging, a Zn anode releases zincate ions (in the form of Zn(OH)$_4^{2-}$ at high alkalinity), which at its saturation decompose into ZnO precipitates. Given the high local zincate concentration, the anode surface is a favored site of the precipitation, which creates a thick oxide layer that limits the transports of electrons and ions and prevents further discharging. A high DoD aggravates passivation as more zincate ions are released to the electrolyte. SCAEs can help by accommodating much more zincate ions. Consistent with previous studies, the solubility increases monotonically from ~0.5 M with $C_{KOH}$ = 6 M to over 1.5 M with $C_{KOH}$ = 15 M (Fig. 4c), measured via an inductively coupled plasma-optical emission spectrophotometer (ICP-OES) after dissolving ZnO powder into the respective electrolytes. Note that the value of $C_{KOH}$ no longer corresponds to the final concentration of either K$^+$ or OH$^-$, as the dissolution consumes both water and OH$^-$.

We further demonstrate the suppression of passivation via chronopotentiometry. In a home-made three-electrode cell (Fig. S19b), we hold a Zn foil anode at 4 mA/cm$^2$ and record the voltage ascending against Ag/AgO. The chronopotentiometric curves (solid lines in Fig. 4d) reflect the equilibrium at the anode surface; they all begin with a graduate rise in voltage as the zincate concentration builds up, continue with a plateau signaling zincate saturation, and end with a sharp rise under passivation. The difference among the electrolytes is how soon the passivation arrives as we expect from the zincate solubility difference. With $C_{KOH}$ = 6 M, passivation occurs after 198 s and the delivery of less than 0.2 mAh/cm$^2$ of areal capacity, whereas with $C_{KOH}$ = 15 M, the delivered areal capacity increases to 3.7 mAh/cm$^2$. The difference persists even if we saturate the electrolytes with zincate before the chronopotentiometry. While one would expect in this case all the zincate-saturated electrolytes to behave similarly and passivate the anode immediately, SCAEs can still deliver much higher areal capacities before passivation (dashed lines in Fig. 4d). We attribute the observation to the dominating role of kinetics instead of thermodynamics. The slow kinetics of passivation, though allowing a high areal capacity, undermines the dimensional stability of the anode, as the zincate ions can be transported far away from the anode. This issue of shape changes can be mitigated with additives such as $Ca(OH)_2$, believed to lower the zincate solubility by forming calcium zincate. We thus measure the zincate solubility in the electrolytes with 10 wt% $Ca(OH)_2$ (in the form of a suspension as it is sparingly soluble). As shown in Fig. 4c, the solubility decreases for all, but not substantially, suggesting the prevailing effect of SACEs in suppressing passivation.

*Super-concentrated alkaline Zn batteries*

We apply SCAEs to Zn‖NiOOH batteries to boost its cycling stability at a high $DoD_{Zn}$. Previous work has suggested that only at a $DoD_{Zn}$ higher than 40% can Zn‖NiOOH batteries compete with commercial lithium-ion batteries in terms of cell-level energy density[23]. At a high DoD, HER and passivation become bigger challenges, which we address with SCAEs. All the electrolytes, either SCAEs and those with lower $C_{KOH}$ as controls, comprise saturated zincate ions and 10 wt% $Ca(OH)_2$. We leverage bi-continuous nanoporous Zn anodes, whose scanning electron microscopic (SEM) images are shown in Figs 5a and S20. The highly connected metal phase has been shown previously by our group to sustain a high conductivity and a uniform reaction distribution necessary for stable cycling at a high $DoD_{Zn}$[27]. An oversized $NiOOH/Ni(OH)_2$ cathode is harvested from a charged commercial battery. They are assembled into a coin cell (Fig. 5a) with lean electrolytes (0.05-0.1 $mL/cm^2$) to ensure the practical relevance of the testing results.

Cells with an SCAE ($C_{KOH}$ = 15 M) outlast those with a regular concentrated alkaline electrolyte ($C_{KOH}$ = 6 M), as shown in Figs 5c and d. All the cells are discharged at a constant current (15 mA/cm$^2$) until reaching a capacity target (determined based on the initial weight of Zn) or a cutoff voltage (1.35 V), and charged at the same current and then a constant voltage (1.9 V) towards the same capacity target. As we do not set a current cutoff for the constant-voltage charging, the cell can always be charged to the capacity target, but not so in discharging, with the loss going to HER in the anode. Therefore, the capacity retention becomes a direct measurement of the amount of the side reaction, although the causes can root in both the loss of Zn mass from the anode and the spontaneous HER. At 40% DoD$_{Zn}$, the SCAE cell retains 80% of the capacity after 190 cycles, in contrast to 45 cycles with $C_{KOH}$ = 6 M (Fig. 5b). The difference is even bigger at 60% DoD$_{Zn}$. The cell with $C_{KOH}$ = 6 M struggles to deliver the capacity target for more than 10 cycles likely due to heavy passivation, whereas that with the SCAE lasts for more than 120 cycles before losing 20% of the capacity (Fig. 5d). While the cycle numbers appear low compared with thousands of cycles achieved in the literature with near-neutral Zn batteries, we highlight the significance of the high cumulative areal capacity (Fig. 5c), as advocated previously by Ma et al. for fulfilling the potential of Zn batteries[62]. For the four cells shown in Figs. 5b and d, we sum the discharging capacities of every cycle until the point of 60% capacity retention. For both electrolytes, cycling at 60% DoD$_{Zn}$ leads to a lower cumulative capacity, consistent with general observations in the literature[63]. Nonetheless, all are much above the range of cumulative capacities in other neutral or weak acid electrolytes surveyed in Ref. 62, which were already overestimates since they were based on symmetric or half cells. The cell with the SCAE ($C_{KOH}$ = 15 M) at 40% DoD$_{Zn}$ achieves 8.4 Ah/cm$^2$, approaching the target of commercial viability of 10 Ah/cm$^2$ adopted from discussions on lithium-metal batteries[64].

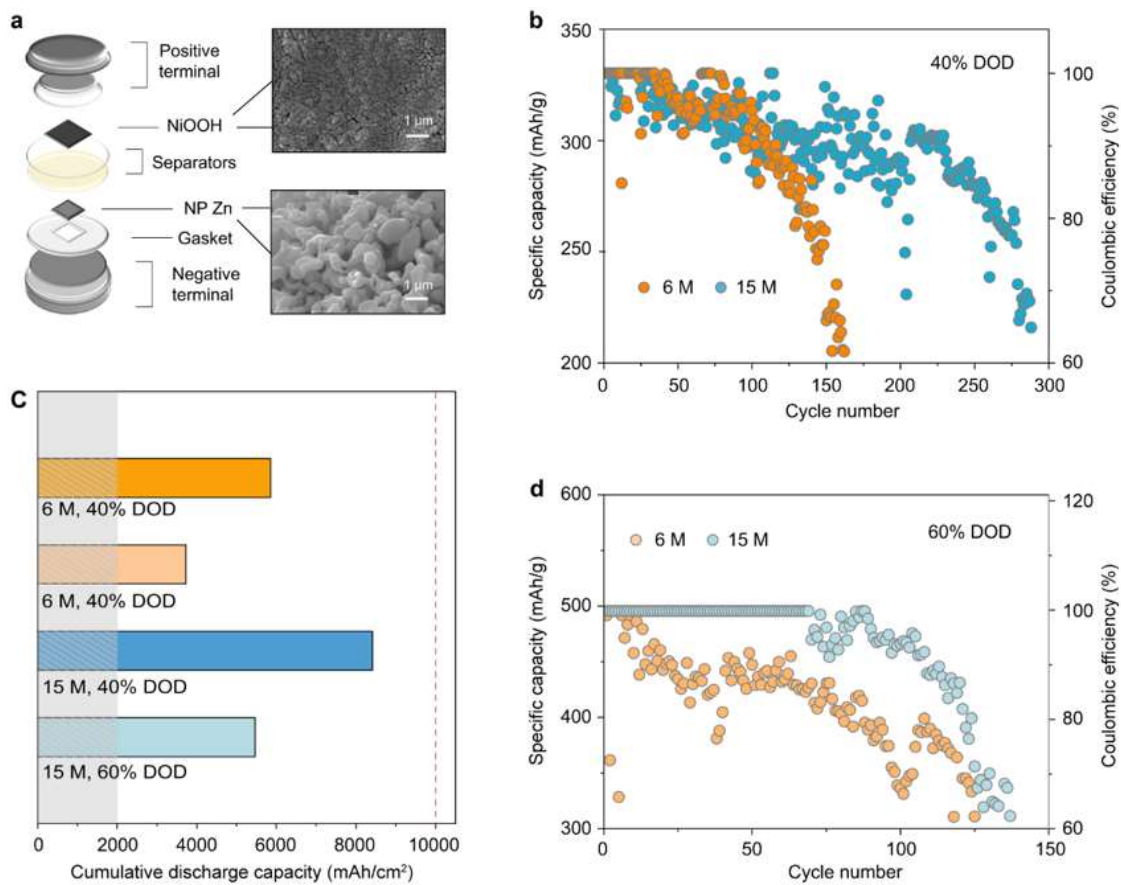

**Fig. 5 | Zn‖NiOOH battery performance. a,** Schematic of the Zn‖NiOOH coin cell with the SEM images of the NiOOH electrode (up) and the NP Zn electrode (down) shown on the right. **b,** Cycling performance of the Zn‖NiOOH cells at 40% DOD. **c,** Cumulative areal capacity plated of the Zn‖NiOOH cells. The dash line shows a commercial target with an ideal coulombic efficiency at 100%, the goal for a lithium metal battery. The shaded area corresponds to the range of cumulative plated capacity compiled by Ref. 62. **d,** Cycling performance of the Zn‖NiOOH coin cells at 60% DOD.

SCAEs can also extend the cycle life of a Zn-air battery. The cell is assembled in a similar build as the earlier Zn‖NiOOH coin cell with a lean electrolyte (0.05 mL/cm$^2$), more relevant to practical applications. Passivation becomes an even bigger issue here as the half-open cell structure and the water-based cathode reaction cause large swings of water content and zincate concentration in the cell. In the control group with $C_{KOH}$ = 6 M, the discharging voltage of the cell quickly descends to the cutoff at about the 20$^{th}$ cycle (Fig. 6a), accompanied by oxide/hydroxide precipitates on both the anode (Zn foil) and the cathode (Pt/C/IrO$_2$ on a carbon paper), confirmed via post-mortem SEM (Fig. S21). The issue lessens when we raise $C_{KOH}$, eventually to 15 M (Figs 6b and c), with which the cell retains more than 60% of the initial capacity at a current of 10 mA/cm$^2$ for more than 110 hours (Fig. 6e), even if the overvoltage (without reaching passivation) in both charging and discharging increases, likely because of the low oxygen solubility in the concentrated electrolytes[65].

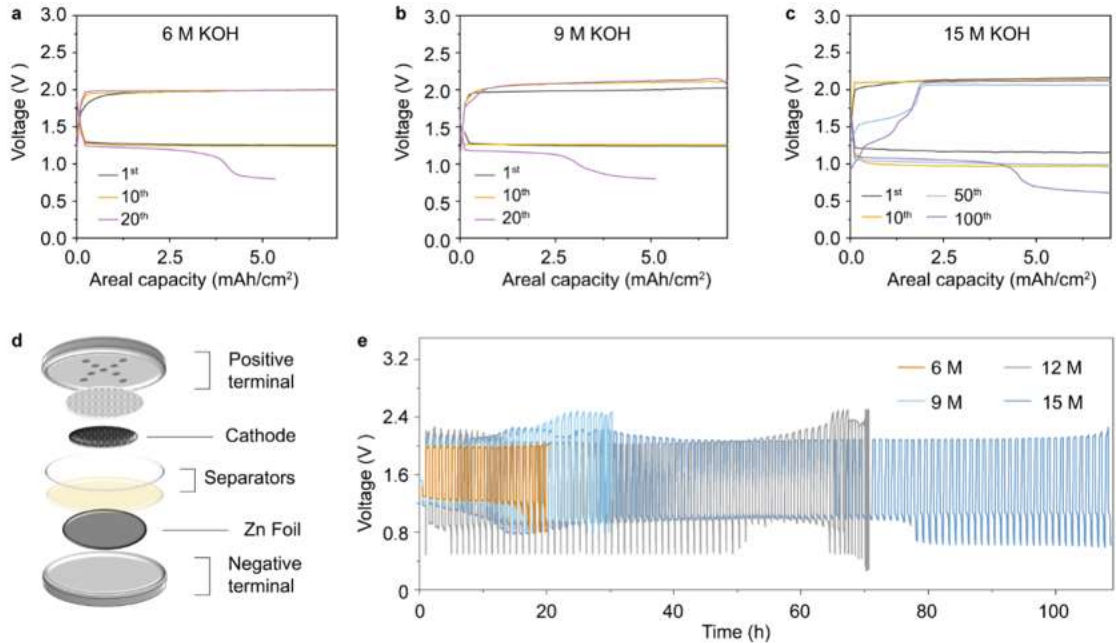

**Fig. 6 | Zn-air battery performances.** Discharging and charging polarization curves of Zn-air coin-cell with a **a**, 6 M **b**, 9 M and **c**, 15 M KOH electrolyte. The 1$^{st}$, 10$^{th}$, 20$^{th}$, and 100$^{th}$ (if attainable) cycles are shown for comparisons. **d**, Schematic of the Zn-air coin-cell. **e**, Voltage vs. time curves at 10 mA/cm$^2$ of the Zn-air cell with electrolytes of the different KOH concentrations.

*Discussion*

The work shows the evolving design principles of alkaline electrolytes for Zn batteries. SCAEs were not favored in the early development largely due to the associated high zincate solubilities that cause large shape changes. Yet these batteries were run at low or moderate DoDs, where passivation and HER were less likely to dictate the cycle lives. Current research on Zn batteries instead strives to increase the DoD to compete with LIBs, which aggravate passivation and HER. Along with new approaches to suppress shape changes (e.g., the nanoporous Zn electrode), we demonstrate that SCAEs can stabilize different types of rechargeable Zn batteries with the large electrochemical window and the slow passivation, while maintaining the rate capability with the high ionic conductivity.

While we focus on KOH, the concept of SCAE can generally apply to other alkali hydroxides to exploit new interactions among species in solutions for both fundamental insights and practical applications. For example, we have observed a high ionic conductivity and stability with SCAEs based on NaOH. Compared with KOH, NaOH has a higher water solubility but a lower dissociation constant, and $Na^+$ is known to complex more strongly with water molecules.[34] It leads to different properties (Fig. S22 and Table S6, S7) and offers potentially more design space for battery applications. Underlying the composition-property relationships are exotic molecular interactions, partly revealed by our spectroscopy and simulation. These alkali hydroxide electrolytes are among the simpler super-concentrated electrolytes, yet they can offer no fewer insights into unique lean-water solution structures, particularly given the well-established experimental and theoretical tools for the investigations[48, 51].

The applications of SCAEs are certainly not limited to what has been demonstrated here. Even for Zn||NiOOH and Zn-air batteries, we expect the current SCAEs to be further optimized given

the tradeoff readily observable in our results. For example, shape changes, potentially aggravated by the high zincate solubility in SCAEs, are currently mitigated by Ca(OH)$_2$ suspensions, which are not designed to function at extreme alkalinity. We expect other additives to better strike the balance between alleviating passivation and suppressing shape changes. We also expect SCAEs to benefit other electrochemical systems for energy storage, conversion, and chemical transformation such as electrochemical synthesis that strives to outcompete water splitting at high efficiencies, where the stability and the conductivity of SCAEs can help.

*Experimental section*

**Electrolyte Preparation**

Electrolytes were prepared by weighing different masses (molality) of commercial high-purity chemical potassium hydroxide (99.99%, metals basis, aladdin) dissolved in ultrapure water and labelled as 1 M, 3 M, 6 M, 9 M, 12 M, 15 M KOH electrolyte. In addition, 1 M and 3 M KCl (99.8%, GR, aladdin), NaOH (99.9%, metals basis, aladdin), and NaCl (99.8%, GR, aladdin), electrolytes were prepared for control experiments. For zinc batteries, we prepared zincate-saturated electrolytes by dissolving ZnO powder (J&K, >300 nm) in KOH electrolytes at 60 ºC, which was then brought back to room temperature and filtered.

**Characterizations**

Nuclear magnetic resonance (NMR) was conducted on Bruker Avance NEO 400 MHz spectrometer. Fourier transform infrared spectroscopy (FTIR) was conducted on Vertex 70 Hyperion 1000 (Bruker). Raman spectroscopy was conducted on a Horiba LabRAM HR Evolution microscope. The surface morphology of Zn anode was characterized by scanning electron microscopy (JEOL-6390 SEM, JEOL-6700 SEM). The Inductively Coupled Plasma-Optical Emission Spectrophotometer (ICP-OES) for solubility test was conducted on Varian, 725-ES. The differential electrochemical mass spectrometry (DEMS) was conducted on Shanghai Linglu Instrument Equipment.

Measurements of electrochemical stability window and hydrogen evolution were performed with linear sweep voltammetry (LSV) on a Pt, Au, Zn or glass carbon electrode at 5 mV/s. Pt foil was used as the counter electrode, and a Hg/HgO electrode was used as the reference electrode. The differential electrochemical mass spectrometry (DEMS) of coin cells (Zn foil as the anode and NiOOH as the cathode) at 25 °C were conducted during the charging-discharging process at a current density of 1 mA cm$^{-2}$. The Zn foil was employed for the anode.

**Battery tests**

Both Zn∥NiOOH and Zn-air batteries were assembled as coin cells. The bi-continuous NP Zn anode was fabricated by reducing ZnO to Zn. We started by coating Sn on the copper foam (1.6 mm thick, purity >99.99%, MTI Corp) at −1 V (vs Ag/AgCl in saturated KCl) with 0.1 M SnCl$_2$ and 0.4 M K$_4$P$_2$O$_7$ aqueous solution. The Sn coated Cu foam was then rinsed with deionized water, acetone (>99.5%) and vacuum dried at room temperature. Commercial ZnO powder (J&K, >300 nm) was pressed onto the Sn coated Cu foam for further reduction. The electrochemical reduction of ZnO to Zn was performed at −1.58 V (vs Ag/AgCl in 1 M KCl) in 3 M KOH until the current plateaued. A Ni foam (1.6 mm thick, purity >99.99%, MTI Corp) was used as counter electrode. After that, the sample was rinsed with deionized water and then sequentially placed in acetone by continuously passing nitrogen gas (>99.5%) for more than 0.5 h to remove H$_2$O and KOH and then vacuum dried at room temperature. The as-fabricated NP Zn was directly used as the battery anode. For the Zn∥NiOOH battery, the cell further comprised a NiOOH/Ni(OH)$_2$ cathode, and porous separators including a nonwoven cellulose membrane (16 mm in diameter, 100 μm thick, 75% porosity) and a Celgard 3501 separator (16 mm diameter, thickness: 25 μm, porosity: 55%). The NiOOH/Ni(OH)$_2$ cathodes (1 cm$^2$) were obtained and directly used from fully charged commercial NiMH AAA batteries (2600 mAh, GP Batteries). The solution used as the electrolyte was different $C_{KOH}$ electrolyte with saturated ZnO for the cathode and separators. The solution consisting of different $C_{KOH}$ electrolyte with saturated ZnO with an additional 10 wt% of Ca(OH)$_2$ was used as the electrolyte for the anode side. The total volume of the electrolyte was controlled to be ~100 μL. All the cells were galvanostatically cycled at a constant current (15 mA/cm$^2$) with a Neware battery testing station (CT-3008W) until reaching a capacity target (40 or 60% of the theoretical capacity determined based on the initial weight of Zn anode) or a cutoff voltage (1.9 V for charging and 1.35 V for discharging). If a cell failed to reach the capacity in charging before reaching the cutoff voltage, a step of constant-voltage charging at 1.9 V would be added to reach the capacity. The Zn-air battery was assembled as an electro-polished Zn foil as the anode (Φ10 mm). The air cathode (1 cm$^2$) was made by mixing equal weights of Pt/C (20 wt% Pt loading, Sigma-Aldrich) and IrO$_2$ (Sigma-Aldrich) into a solution of DI water, Nafion, isopropanol to form a suspension that was dispersed on carbon cloth (CeTech Co., Ltd, W1S1005, 0.785 cm$^2$). The loading of the catalyst is about 5 mg/cm$^2$. Ni foam (1.6 mm thick, purity >99.99%, MTI Corp) was used as the cathode current collector. The electrolyte was a KOH solution. Galvanostatic cycling was performed inside a sealed climatic chamber (filled with humidified O$_2$) at a current density of 10 mA/cm$^2$, with 30 mins for each discharge/charge. All the electrochemical measurements are carried out at a room temperature near 25 °C.

**Computational methods**

We conducted ab initio molecular dynamics (AIMD) simulations within the Born–Oppenheimer approximation using the Qbox code, v.1.72.2[66], with a time step of 0.24 fs. We used the SCAN[67] exchange-correlation functional and the SG15 ONCV norm-conserving pseudopotentials[68, 69], with a kinetic energy cut-off of 65 Ry for the plane wave basis set. The simulation box and the number of $H_2O$ and KOH molecules in the cubic supercell with periodic boundary conditions are detailed in Table S5. The temperature was regulated using the BDP[70] thermostat with stochastic velocity rescaling (τ = 24.2 fs), enabling the generation of a canonical ensemble (NVT) at 330K. Trajectories were collected for ∼45 ps at each concentration, after initial equilibration of ∼5 ps. Forces on the nuclei are calculated on the fly from the instantaneous ground state of electrons using density functional theory (DFT), and charge transfer and polarization, critical in electrochemistry, are both considered at the same theory level without any empirical parameters or experimental inputs.

## *References*

1.  Borodin, O., Self, J., Persson, K.A., Wang, C. & Xu, K. Uncharted Waters: Super-Concentrated Electrolytes. *Joule* **4**, 69-100 (2020).
2.  Sayah, S. et al. How do super concentrated electrolytes push the Li-ion batteries and supercapacitors beyond their thermodynamic and electrochemical limits? *Nano Energy* **98**, 107336 (2022).
3.  Suo, L. et al. "Water-in-salt" electrolyte enables high-voltage aqueous lithium-ion chemistries. *Science* **350**, 938-943 (2015).
4.  Yamada, Y. et al. Hydrate-melt electrolytes for high-energy-density aqueous batteries. *Nature Energy* **1**, 16129 (2016).
5.  Xie, J., Liang, Z. & Lu, Y.-C. Molecular crowding electrolytes for high-voltage aqueous batteries. *Nature Materials* **19**, 1006-1011 (2020).
6.  Suo, L. et al. Advanced High-Voltage Aqueous Lithium-Ion Battery Enabled by "Water-in-Bisalt" Electrolyte. *Angewandte Chemie International Edition* **55**, 7136-7141 (2016).

7. Wang, F. et al. Highly reversible zinc metal anode for aqueous batteries. *Nature Materials* **17**, 543-549 (2018).

8. Zhang, C. et al. A ZnCl2 water-in-salt electrolyte for a reversible Zn metal anode. *Chemical Communications* **54**, 14097-14099 (2018).

9. Zhang, Q. et al. Chaotropic Anion and Fast-Kinetics Cathode Enabling Low-Temperature Aqueous Zn Batteries. *ACS Energy Letters* **6**, 2704-2712 (2021).

10. Zhang, Q. et al. Designing Anion-Type Water-Free Zn2+ Solvation Structure for Robust Zn Metal Anode. *Angewandte Chemie International Edition* **60**, 23357-23364 (2021).

11. Ma, Y. et al. N,N-dimethylformamide tailors solvent effect to boost Zn anode reversibility in aqueous electrolyte. *National Science Review* **9**, nwac051 (2022).

12. Tang, B., Shan, L., Liang, S. & Zhou, J. Issues and opportunities facing aqueous zinc-ion batteries. *Energy & Environmental Science* **12**, 3288-3304 (2019).

13. Guo, S. et al. Fundamentals and perspectives of electrolyte additives for aqueous zinc-ion batteries. *Energy Storage Materials* **34**, 545-562 (2021).

14. Yuan, L. et al. Regulation methods for the Zn/electrolyte interphase and the effectiveness evaluation in aqueous Zn-ion batteries. *Energy & Environmental Science* **14**, 5669-5689 (2021).

15. Hoang, D. et al. Vanillin: An Effective Additive to Improve the Longevity of Zn Metal Anode in a 30 m ZnCl2 Electrolyte. *Advanced Energy Materials* **13**, 2301712 (2023).

16. Zhang, L. et al. ZnCl2 "Water-in-Salt" Electrolyte Transforms the Performance of Vanadium Oxide as a Zn Battery Cathode. *Advanced Functional Materials* **29**, 1902653 (2019).

17. Tang, X. et al. Unveiling the Reversibility and Stability Origin of the Aqueous V2O5–Zn Batteries with a ZnCl2 "Water-in-Salt" Electrolyte. *Advanced Science* **8**, 2102053 (2021).

18. Zhang, Y. et al. Water or Anion? Uncovering the Zn2+ Solvation Environment in Mixed Zn(TFSI)2 and LiTFSI Water-in-Salt Electrolytes. *ACS Energy Letters* **6**, 3458-3463 (2021).

19. Olbasa, B.W. et al. Highly Reversible Zn Metal Anode Stabilized by Dense and Anion-Derived Passivation Layer Obtained from Concentrated Hybrid Aqueous Electrolyte. *Advanced Functional Materials* **32**, 2103959 (2022).

20. Dong, D., Wang, T., Sun, Y., Fan, J. & Lu, Y.-C. Hydrotropic solubilization of zinc acetates for sustainable aqueous battery electrolytes. *Nature Sustainability* **6**, 1474-1484 (2023).

21. Lohmann, K.B. et al. Solvent-Free Method for Defect Reduction and Improved Performance of p-i-n Vapor-Deposited Perovskite Solar Cells. *ACS Energy Letters* **7**, 1903-1911 (2022).

22. Shang, W. et al. Rechargeable alkaline zinc batteries: Progress and challenges. *Energy Storage Materials* **31**, 44-57 (2020).

23. Parker, J.F. et al. Rechargeable nickel–3D zinc batteries: An energy-dense, safer alternative to lithium-ion. *Science* **356**, 415-418 (2017).

24. Adler, T.C., McLarnon, F.R. & Cairns, E.J. Investigations of a New Family of Alkaline−Fluoride−Carbonate Electrolytes for Zinc/Nickel Oxide Cells. *Industrial & Engineering Chemistry Research* **37**, 3237-3241 (1998).

25. Bass, K., Mitchell, P.J., Wilcox, G.D. & Smith, J. Methods for the reduction of shape change and dendritic growth in zinc-based secondary cells. *Journal of Power Sources* **35**, 333-351 (1991).

26. Parker, J.F., Pala, I.R., Chervin, C.N., Long, J.W. & Rolison, D.R. Minimizing Shape Change at Zn Sponge Anodes in Rechargeable Ni–Zn Cells: Impact of Electrolyte Formulation. *Journal of The Electrochemical Society* **163**, A351 (2016).

27. Lim, M.B., Lambert, T.N. & Ruiz, E.I. Effect of ZnO-Saturated Electrolyte on Rechargeable Alkaline Zinc Batteries at Increased Depth-of-Discharge. *Journal of The Electrochemical Society* **167**, 060508 (2020).

28. Li, L. et al. Phase-transition tailored nanoporous zinc metal electrodes for rechargeable alkaline zinc-nickel oxide hydroxide and zinc-air batteries. *Nature Communications* **13**, 2870 (2022).

29. Chen, S. et al. Aqueous rechargeable zinc air batteries operated at −110°C. *Chem* **9**, 497-510 (2023).

30. Droguet, L., Grimaud, A., Fontaine, O. & Tarascon, J.-M. Water-in-Salt Electrolyte (WiSE) for Aqueous Batteries: A Long Way to Practicality. *Advanced Energy Materials* **10**, 2002440 (2020).

31. Agmon, N. et al. Protons and Hydroxide Ions in Aqueous Systems. *Chemical Reviews* **116**, 7642-7672 (2016).

32. Dahms, F., Fingerhut, B.P., Nibbering, E.T.J., Pines, E. & Elsaesser, T. Large-amplitude transfer motion of hydrated excess protons mapped by ultrafast 2D IR spectroscopy. *Science* **357**, 491-495 (2017).

33. Aziz, E.F., Ottosson, N., Faubel, M., Hertel, I.V. & Winter, B. Interaction between liquid water and hydroxide revealed by core-hole de-excitation. *Nature* **455**, 89-91 (2008).

34. Tuckerman, M.E., Chandra, A. & Marx, D. Structure and Dynamics of OH-(aq). *Accounts of Chemical Research* **39**, 151-158 (2006).

35. Yang, C. et al. All-temperature zinc batteries with high-entropy aqueous electrolyte. *Nature Sustainability* **6**, 325-335 (2023).

36. Lv, T. & Suo, L. Water-in-salt widens the electrochemical stability window: Thermodynamic and kinetic factors. *Current Opinion in Electrochemistry* **29**, 100818 (2021).

37. Han, J., Mariani, A., Passerini, S. & Varzi, A. A perspective on the role of anions in highly concentrated aqueous electrolytes. *Energy & Environmental Science* **16**, 1480-1501 (2023).

38. Seidell, A. & Linke, W.F. Solubilities of Inorganic and Organic Compounds: A Compilation of Solubility Data from the Periodical Literature. (Van Nostrand, 1952).

39. Goyal, A. & Koper, M.T.M. The Interrelated Effect of Cations and Electrolyte pH on the Hydrogen Evolution Reaction on Gold Electrodes in Alkaline Media. *Angewandte Chemie International Edition* **60**, 13452-13462 (2021).

40. Schmidt, T.J., Ross, P.N. & Markovic, N.M. Temperature dependent surface electrochemistry on Pt single crystals in alkaline electrolytes: Part 2. The hydrogen evolution/oxidation reaction. *Journal of Electroanalytical Chemistry* **524-525**, 252-260 (2002).

41. Gilliam, R.J., Graydon, J.W., Kirk, D.W. & Thorpe, S.J. A review of specific conductivities of potassium hydroxide solutions for various concentrations and temperatures. *International Journal of Hydrogen Energy* **32**, 359-364 (2007).

42. Harris, K.R. On the Use of the Angell–Walden Equation To Determine the "Ionicity" of Molten Salts and Ionic Liquids. *The Journal of Physical Chemistry B* **123**, 7014-7023 (2019).

43. Licht, S. pH Measurement in Concentrated Alkaline Solutions. *Analytical Chemistry* **57**, 514-519 (1985).

44. Drexler, C.I. et al. Counter Cations Affect Transport in Aqueous Hydroxide Solutions with Ion Specificity. *Journal of the American Chemical Society* **141**, 6930-6936 (2019).

45. Walrafen, G.E. & Douglas, R.T.W. Raman spectra from very concentrated aqueous NaOH and from wet and dry, solid, and anhydrous molten, LiOH, NaOH, and KOH. *The Journal of Chemical Physics* **124**, 114504 (2006).

46. Zhang, Q. et al. Modulating electrolyte structure for ultralow temperature aqueous zinc batteries. *Nature Communications* **11**, 4463 (2020).

47. Stefanski, J., Schmidt, C. & Jahn, S. Aqueous sodium hydroxide (NaOH) solutions at high pressure and temperature: insights from in situ Raman spectroscopy and ab initio molecular dynamics simulations. *Physical Chemistry Chemical Physics* **20**, 21629-21639 (2018).

48. Imberti, S. et al. Ions in water: The microscopic structure of concentrated hydroxide solutions. *The Journal of Chemical Physics* **122**, 194509 (2005).

49. Luzar, A. & Chandler, D. Hydrogen-bond kinetics in liquid water. *Nature* **379**, 55-57 (1996).

50. Ikeda, T., Boero, M. & Terakura, K. Hydration of alkali ions from first principles molecular dynamics revisited. *The Journal of Chemical Physics* **126**, 034501 (2007).

51. Zhu, Z. & Tuckerman, M.E. Ab Initio Molecular Dynamics Investigation of the Concentration Dependence of Charged Defect Transport in Basic Solutions via Calculation of the Infrared Spectrum. *The Journal of Physical Chemistry B* **106**, 8009-8018 (2002).

52. Chen, B., Ivanov, I., Park, J.M., Parrinello, M. & Klein, M.L. Solvation Structure and Mobility Mechanism of OH-: A Car−Parrinello Molecular Dynamics Investigation of Alkaline Solutions. *The Journal of Physical Chemistry B* **106**, 12006-12016 (2002).

53. Sbroscia, M., Sodo, A., Bruni, F., Corridoni, T. & Ricci, M.A. OH Stretching Dynamics in Hydroxide Aqueous Solutions. *The Journal of Physical Chemistry B* **122**, 4077-4082 (2018).

54. LaCount, M.D. & Gygi, F. Ensemble first-principles molecular dynamics simulations of water using the SCAN meta-GGA density functional. *The Journal of Chemical Physics* **151**, 164101 (2019).

55. Lileev, A.S., Loginova, D.V. & Lyashchenko, A.K. Microwave dielectric properties of potassium hydroxide aqueous solutions. *Russian Journal of Inorganic Chemistry* **56**, 961-967 (2011).

56. Shao, Y. et al. Temperature effects on the ionic conductivity in concentrated alkaline electrolyte solutions. *Physical Chemistry Chemical Physics* **22**, 10426-10430 (2020).

57. Marx, D., Chandra, A. & Tuckerman, M.E. Aqueous Basic Solutions: Hydroxide Solvation, Structural Diffusion, and Comparison to the Hydrated Proton. *Chemical Reviews* **110**, 2174-2216 (2010).

58. Tuckerman, M.E., Marx, D. & Parrinello, M. The nature and transport mechanism of hydrated hydroxide ions in aqueous solution. *Nature* **417**, 925-929 (2002).

59. Lin, I.C., Seitsonen, A.P., Tavernelli, I. & Rothlisberger, U. Structure and Dynamics of Liquid Water from ab Initio Molecular Dynamics—Comparison of BLYP, PBE, and revPBE Density Functionals with and without van der Waals Corrections. *Journal of Chemical Theory and Computation* **8**, 3902-3910 (2012).

60. Chen, M. et al. Hydroxide diffuses slower than hydronium in water because its solvated structure inhibits correlated proton transfer. *Nature Chemistry* **10**, 413-419 (2018).

61. Codorniu-Hernández, E. & Kusalik, P.G. Probing the mechanisms of proton transfer in liquid water. *Proceedings of the National Academy of Sciences* **110**, 13697-13698 (2013).

62. Ma, L. et al. Realizing high zinc reversibility in rechargeable batteries. *Nature Energy* **5**, 743-749 (2020).

63. Turney, D.E. et al. Rechargeable Zinc Alkaline Anodes for Long-Cycle Energy Storage. *Chemistry of Materials* **29**, 4819-4832 (2017).

64. Albertus, P., Babinec, S., Litzelman, S. & Newman, A. Status and challenges in enabling the lithium metal electrode for high-energy and low-cost rechargeable batteries. *Nature Energy* **3**, 16-21 (2018).

65. Zhang, C., Fan, F.-R.F. & Bard, A.J. Electrochemistry of Oxygen in Concentrated NaOH Solutions: Solubility, Diffusion Coefficients, and Superoxide Formation. *Journal of the American Chemical Society* **131**, 177-181 (2009).

66. Gygi, F. Architecture of Qbox: A scalable first-principles molecular dynamics code. *IBM Journal of Research and Development* **52**, 137-144 (2008).

67. Sun, J., Ruzsinszky, A. & Perdew, J.P. Strongly Constrained and Appropriately Normed Semilocal Density Functional. *Physical Review Letters* **115**, 036402 (2015).

68. Hamann, D.R. Optimized norm-conserving Vanderbilt pseudopotentials. *Physical Review B* **88**, 085117 (2013).

69. Schlipf, M. & Gygi, F. Optimization algorithm for the generation of ONCV pseudopotentials. *Computer Physics Communications* **196**, 36-44 (2015).

70. Bussi, G., Donadio, D. & Parrinello, M. Canonical sampling through velocity rescaling. *The Journal of Chemical Physics* **126**, 014101 (2007).

*Acknowledgment*

We acknowledge the funding support from the National Foundation of Natural Science, China (52022002), and the Research Grant Council, Hong Kong (C1002-21G). D. P. acknowledges support from the Research Grant Council, Hong Kong (16306621), the Croucher Foundation through the Croucher Innovation Award, the National Natural Science Foundation, China (22022310), and the Hetao Shenzhen/Hong Kong Innovation and Technology Cooperation (HZQB-KCZYB-2020083).

*Author contributions*

Q.C. and P.D. supervised the research. Y.M. and J.H. designed and performed the experiments and the simulation. Y.M. and S.G. performed the characterizations. Y.M. and L.L. carried out the battery tests. L.L. and Z.Y. advised the tests of the Zn-air batteries. C.K.K.C. and D.X. prepared the anode materials. Q.C., P.D., Y.M. and J.H. wrote the paper and all the authors discussed it.

*Competing interests*

The authors declare no competing interests.